# Monitoring exposure-length variations in submarine power cables using distributed fiber-optic sensing

**Sakiko Mishima[1][*], Yoshiyuki Yajima[1], Noriyuki Tonami[1], Tomoyuki Hino[1], Shugo Aibe[1], Junichiro Saikawa[1] and Koji Mizuguchi[1]**

[1] NEC Corporation, Tokyo, Japan

[*] E-mail: s.mishima@nec.com

**Abstract.** This study proposes an anomaly-detection framework for monitoring exposure-length variations in submarine free-span cables using Distributed Acoustic Sensing (DAS), which is one of the distributed fiber-optic sensing technologies. To address environmental variability and limited training data in offshore environments, a regression-based feature extraction method was introduced to derive low-dimensional latent representations that retain exposure-length-dependent vibration characteristics while suppressing environmental influences. The extracted features were used for one-class Support Vector Machine (SVM)-based anomaly detection. The proposed framework was evaluated through wave-tank experiments with exposure lengths ranging from 2 to 10 m. Experimental results showed that anomaly scores decreased approximately monotonically with increasing exposure-length change, exhibiting a strong correlation ($r = -0.83$). The binary classification achieved an F1 score of 0.82 despite training with only small-sample datasets. These findings demonstrate that exposure-length variations can be reliably detected under severe data limitations, supporting the potential of DAS-based cable condition monitoring.

## 1. Introduction

Submarine power cables constitute critical infrastructure in offshore wind power generation, as they transmit electricity from offshore turbines to onshore substations. Recent reports indicate that cable failures account for up to 80% of the total financial losses in offshore wind farms [1]. This highlights the substantial economic impact of cable-related faults and underscores the importance of continuous cable condition monitoring to ensure stable and reliable operation. To mitigate fatigue failure caused by external loading and hydrodynamic forces, submarine power cables are typically buried beneath the seabed.

Structural constraints near fixed-bottom turbine foundations often prevent sufficient burial depth. The exposed cable sections formed in these areas are referred to as free spans. These free spans experience elevated mechanical stress, and the risk of damage increases with exposure length [2]. Early detection of exposure-length variations is therefore essential for effective asset integrity management.

The condition of submarine power cables has traditionally been assessed using vessel-based inspection methods, including surveys with Remotely Operated Vehicles (ROVs) and multi-beam echo sounders. However, these technologies face challenges in providing detailed geometric and visual information, as they are costly and weather-dependent, making them unsuitable for continuous, wide-area monitoring. This limitation has motivated growing interest in distributed fiber-optic sensing technologies that use optical communication fibers integrated within submarine power cables as a cost-effective alternative for large-scale monitoring.

Distributed fiber-optic sensing exploits variations in backscattered light propagating along optical fibers that arise from changes in temperature and/or strain. Depending on the measured physical quantity, such techniques are categorized as Distributed Temperature Sensing (DTS) or Distributed Acoustic Sensing (DAS) [3]. By connecting optical interrogators to existing fiber-optic cables, the entire cable route can be converted into a distributed sensor, enabling continuous monitoring over long distances with minimal additional sensing devices.

DTS has been used to estimate the burial state or depth of submarine power cables [4, 5]. However, temperature-based measurements are affected by thermal diffusion, which spatially smooths temperature distributions and limits the detection of localized exposure, such as free spans [6]. In addition, thermal responses are influenced by load variations and environmental boundary conditions, including convective heat transfer, which can attenuate or delay temperature changes associated with minor exposure variations [7]. These factors restrict the reliability of DTS for monitoring free-span evolution.

In recent years, DAS has garnered significant attention as a technology capable of measuring vibrations imparted to high-density submarine power cables over long distances. Figure 1 illustrates typical environmental vibrations acting on submarine cables and the cable condition changes that can potentially be monitored using DAS. There is considerable interest in using DAS for early risk detection, such as identifying anomalies that might cause cable failures. One application uses DAS's ability to monitor specific localized points, enabling the detection of localized structural changes in submarine cables [8, 9].

Although DAS has potential for localized condition monitoring, its offshore deployment faces two main issues: environmental variability and limited data availability. Environmental factors such as wave loading, tidal currents, and ambient noise significantly affect measurements, making it difficult to identify structural signals. Additionally, collecting sufficient abnormal data for supervised anomaly detection is difficult in marine environments. To overcome these problems, Duthé et al. have proposed a self-supervised contrastive learning method [8]. Nonetheless, their approach still requires a large amount of training data, so the challenge of data collection remains only partly resolved.

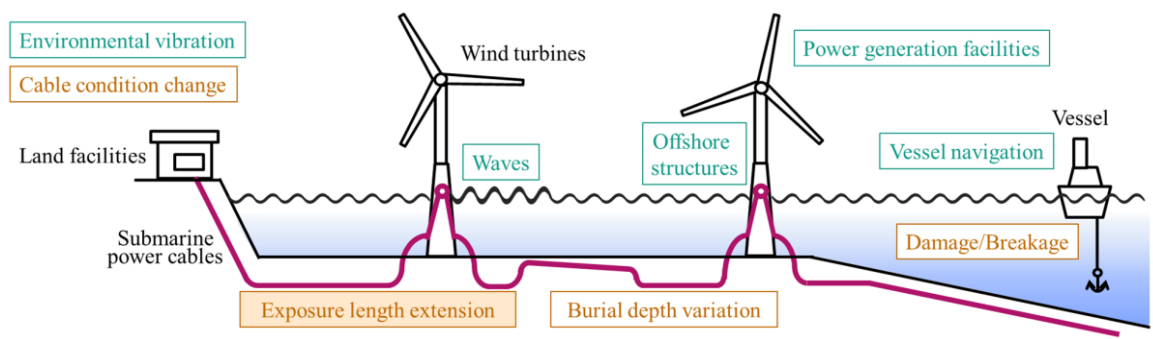

**Figure 1.** Conceptual illustration of environmental vibrations acting on a submarine power cable and the cable condition changes that can be monitored using Distributed Acoustic Sensing (DAS). Wave-induced loads and other environmental excitations induce vibration responses along the cable, with the exposed free-span sections particularly susceptible.

To overcome these practical limitations, this study proposes an anomaly-detection framework for monitoring exposure-length variations in submarine free-span cables using DAS measurements. The proposed method uses intermediate representations derived from a regression model. By exploiting differences in installed exposure lengths, the method extracts low-dimensional latent features that capture exposure-related characteristics from small-sample datasets collected under different installed lengths. Compressing the measured DAS signal into a low-dimensional representation is also expected to enhance robustness against environmental variability. The resulting latent features are subsequently used for anomaly detection. Designed to operate effectively with limited training data, the proposed framework is evaluated in this study through wave-tank experiments under controlled conditions as an initial step toward offshore cable condition monitoring.

## 2. Monitoring challenges in submarine free-span cables

In offshore wind farms with fixed-bottom turbines, foundations are designed according to local bathymetry and environmental loads, resulting in free-span sections with varying exposure lengths within a single wind farm. This structural diversity enables analysis of how cable vibration characteristics depend on exposure length when spatially distributed DAS measurements are available.

DAS captures distributed vibration responses along optical fibers embedded in submarine cables, reflecting both structural dynamics and time-varying environmental excitation such as wave-induced loading and currents. Although exposure-length variations should produce systematic differences in vibration behavior, practical offshore monitoring imposes two constraints. First, environmental excitation varies temporally and spatially, inducing non-stationarity and section-wise statistical differences in DAS signals even without structural changes. Second, monitoring often begins immediately after installation, when only limited normal-condition data are available, and no labeled abnormal samples are available. Consequently, exposure-length variations cannot be reliably identified through direct comparison of raw vibration signals. Moreover, data-intensive learning strategies that require large labeled datasets or extensive pretraining are impractical under such operational conditions.

These considerations imply that an effective monitoring strategy must (i) extract structural information from limited normal-condition data, (ii) suppress the influence of environmental variability, and (iii) operate independently across free-span sections with different installed lengths. Accordingly, this study formulates the task as detecting deviations from section-wise normal behavior associated with changes in free-span exposure length under non-stationary environmental conditions.

## 3. Methodology

To satisfy the monitoring requirements outlined in Section 2, our new monitoring framework has been developed according to the following approach:
- (i) Data-efficient modeling: The method uses a signal-processing or machine-learning approach that can operate with limited normal-condition data, avoiding reliance on large labeled datasets.
- (ii) Exposure-sensitive latent representation: DAS signals are projected into a low-dimensional space that preserves exposure-length-dependent components while suppressing environmentally induced fluctuations.

(iii) Section-wise anomaly detection: Because installation conditions differ across free-span sections, anomaly detection is performed independently for each section. Anomalies are defined as deviations from the baseline normal distribution established at the initial monitoring stage.

We propose an anomaly-detection framework for monitoring exposure-length variations in submarine free-span cables using DAS measurements. Figure 2 shows an overview of the proposed framework. This proposed method employs a combination of regression-based feature extraction and single-class anomaly modelling to achieve robust installation length monitoring even under data-scarce and non-stationary oceanic conditions. In this paper, we employed Partial Least Squares regression (PLS regression) for the feature extraction model and a one-class Support Vector Machine (one-class SVM) for the anomaly detection model, both of which are capable of learning even with a small dataset. Vibration signals from free-span sections with varying installed exposure lengths are used to train a regression model that extracts latent features correlated with exposure length. These features form a low-dimensional representation that emphasizes structural variance while reducing environmental influence. Anomaly detection is then performed in this latent space using anomaly-detection models trained only on normal-condition data to define section-specific decision boundaries.

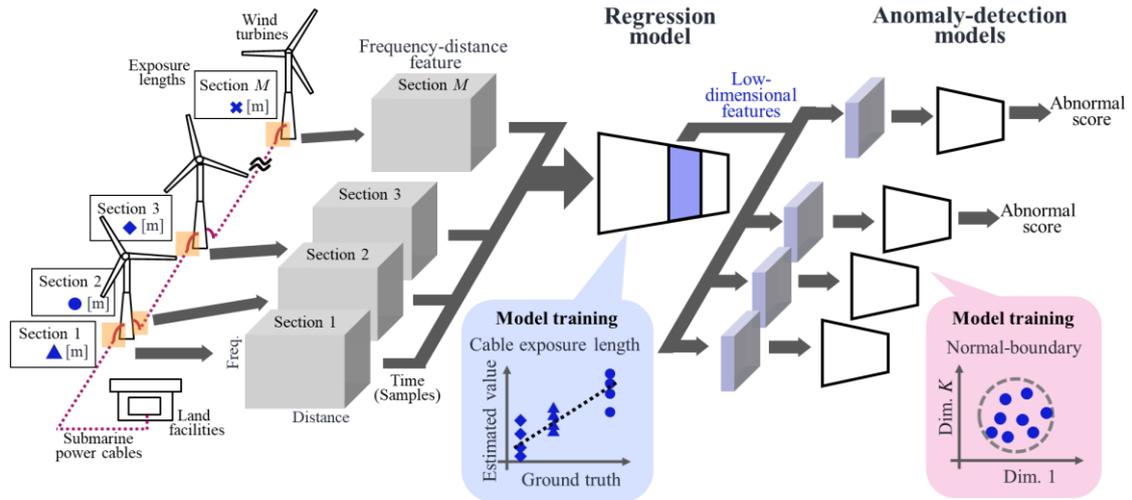

**Figure 2.** Overview of the proposed two-stage monitoring framework combining regression-based feature extraction and anomaly detection for monitoring exposure-length variations in submarine free-span cables using DAS signals.

3.1 Low-dimensional feature extraction using a regression model

DAS signals contain not only structural vibration responses, but also environmental noise induced by waves and currents. To reliably detect anomalies caused by changes in cable exposure length, it is therefore necessary to suppress noise while preserving exposure-length-dependent characteristics. The proposed framework introduces a regression-based feature-extraction step that projects high-dimensional frequency-space DAS signals into a latent space. The regression model is trained with exposure length as the target variable, encouraging the latent representation to retain information relevant to free-span conditions while reducing unrelated variability [10]. The resulting intermediate features thus encode exposure-length-dependent vibration characteristics and are used as inputs to the anomaly-detection model. In this study,

Partial Least Squares (PLS) regression [11] is employed, as it enables dimensionality reduction while explicitly maximizing the relationship between input features and exposure length.

As a preprocessing step, the DAS signal at each spatial point is transformed into a time–frequency representation by applying a Short-Time Fourier Transform (STFT) along the temporal axis. For each time window, the frequency components obtained from multiple spatial locations are concatenated along the spatial dimension to construct a frequency–distance feature representation.

The resulting structured data are arranged into an input matrix $\boldsymbol{X} \in \mathbb{R}^{N \times D}$, where $N$ is the number of time windows (samples) and $D$ represents the total number of frequency–distance bins. The corresponding exposure lengths of the free-span sections are denoted by $\boldsymbol{y} \in \mathbb{R}^{N \times 1}$. These exposure lengths are obtained from independent sources, such as cable installation drawings or inspection records, rather than from DAS measurements. In PLS regression, the relationship between $\boldsymbol{X}$ and $\boldsymbol{y}$ is modeled using a set of latent variable $\boldsymbol{T} \in \mathbb{R}^{N \times K}$, according to

$$\boldsymbol{X} = \boldsymbol{T}\boldsymbol{P}^\top + \boldsymbol{E}, \quad \boldsymbol{y} = \boldsymbol{T}\boldsymbol{q} + \boldsymbol{e}, \quad (1)$$

where $K$ is the number of latent components in the PLS model, which also determines the dimensionality of the latent feature space. $\boldsymbol{P} \in \mathbb{R}^{D \times K}$ and $\boldsymbol{q} \in \mathbb{R}^{K \times 1}$ are the loading matrix and loading vector for the input and output, respectively, and $\boldsymbol{E}$ and $\boldsymbol{e}$ represent residuals. Each latent variable $\boldsymbol{t}_k$ is obtained by projecting the input data onto a weight vector $\boldsymbol{w}_k$, as

$$\boldsymbol{t}_k = \boldsymbol{X}\boldsymbol{w}_k. \quad (2)$$

The weight vectors $\boldsymbol{w}_k$ are determined such that the covariance between the projected input and the output is maximized:

$$\max_{\boldsymbol{w}_k} \mathrm{Cov}(\boldsymbol{X}\boldsymbol{w}_k, \boldsymbol{y}), \quad \text{s.t.} \quad \|\boldsymbol{w}_k\| = 1. \quad (3)$$

This optimization ensures that the extracted latent variables capture the directions in the input space that are most strongly correlated with exposure length.

By using PLS regression as a feature-extraction mechanism, multicollinearity inherent in DAS signals can be effectively mitigated while obtaining low-dimensional latent representations that explicitly reflect exposure-length differences. These latent features form the basis for the subsequent anomaly detection stage described in the following section.

3.2 Anomaly detection using support vector machines

Based on the low-dimensional latent features obtained in the feature extraction step, anomalies are detected as deviations from normal cable behavior using a one-class SVM. The one-class SVM is an unsupervised anomaly detection method that learns the distribution of normal data and determines whether newly observed data deviates from this distribution [12]. This property makes it well-suited for practical monitoring scenarios in which labeled data corresponding to abnormal conditions is unavailable.

In the proposed framework, the initial exposure state of each free-span section is defined as the normal condition. Accordingly, an independent one-class SVM model is constructed for each free-span section using latent features extracted from DAS measurements collected during the initial monitoring period. This section-wise modeling strategy accounts for differences in environmental variability and baseline vibration characteristics among individual free spans.

Let the set of latent feature vectors obtained under normal conditions be denoted by

$$\{\mathbf{t}_i\}_{i=1}^n \in \mathbb{R}^K, \quad (4)$$

where $n$ is the number of training sample, and $K$ is the latent dimensionality defined in Section 3.1. The one-class SVM learns a decision boundary that encloses the region of the latent space occupied by normal data by solving the following optimization problem:

$$\min_{\boldsymbol{\beta},\rho,\xi_i} \left( \frac{1}{2} \| \boldsymbol{\beta} \|^2 + \frac{1}{vn} \sum_{i=1}^{n} \xi_i - \rho \right), \quad \text{s.t.} \quad \boldsymbol{\beta}^\top \phi(\mathbf{t}_i) \geq \rho - \xi_i, \xi_i \geq 0, \quad (5)$$

where $\boldsymbol{\beta}$ is the normal vector of the separating hyperplane, $\rho$ is the offset determining the decision boundary, and $\xi_i$ are slack variables that allow for soft boundary violations. The parameter $v \in (0, 1]$ controls the fraction of support vectors and the upper bound on the fraction of training samples allowed to lie outside the learned boundary. The mapping $\phi(\cdot)$ represents a nonlinear transformation induced by a kernel function; in this study, a Gaussian (radial basis function) kernel is employed.

After training, an anomaly score for a test sample, $\mathbf{t}$ is computed using the decision function

$$f(\mathbf{t}) = \boldsymbol{\beta}^\top \phi(\mathbf{t}) - \rho. \quad (6)$$

Samples for which $f(\mathbf{t}) < 0$ are classified as anomalous, indicating a deviation from normal behavior. In the proposed framework, this anomaly score is evaluated independently for each free-span section, enabling localized detection of exposure-length variations while remaining robust to non-stationary environmental conditions.

## 4. Experimental evaluation

To evaluate the anomaly-detection performance of the proposed method, vibration signals were measured using a power cable with integrated optical fibers installed in a wave tank under different free-span lengths. The purpose of this wave-tank experiment was to provide an initial validation of the proposed framework under controlled and repeatable conditions, rather than to reproduce the full range of environmental conditions encountered in actual offshore sites. The acquired data were split into training and evaluation datasets, and anomaly-detection accuracy was calculated using the evaluation data.

### 4.1 Conditions for signal measurement in wave-tank environments

Vibration measurements were conducted using a DAS system connected to a power cable with integrated optical fibers configured to simulate free-span conditions. The experimental setup is illustrated in Fig. 3. Figure 3(a) shows an overview of the cable fixture body assembled on land. The wave tank used in this study measured 50 m × 8 m × 4.5 m, and the cable fixture was designed to fit the tank dimensions. As shown in Fig. 3(b), the fixture was submerged in the tank and fixed in place with wires to prevent movement during wave excitation. The cable, with a diameter of 43 mm and a weight of 4 kg/m, integrates both power and communication lines, similar to offshore deployment configurations. The integrated optical fiber consists of four cores, two of which were used for measurement. To simulate free-span conditions, a section of the cable was intentionally left unsupported, as shown schematically in Fig. 3(c). This unsupported section, hereafter referred to as the exposed section, was adjusted to lengths from 2 m to 10 m by modifying the bottom supports. Outside the exposed section, the cable was secured to the fixture at regular intervals to ensure structural stability. The fixture was installed in a wave tank equipped with a wave generator (Fig. 3(d)), and the cable axis was oriented at 20° relative to the wave-propagation direction. The 20° configuration was chosen as a representative condition to generate stable and repeatable vibration responses under controlled wave excitation.

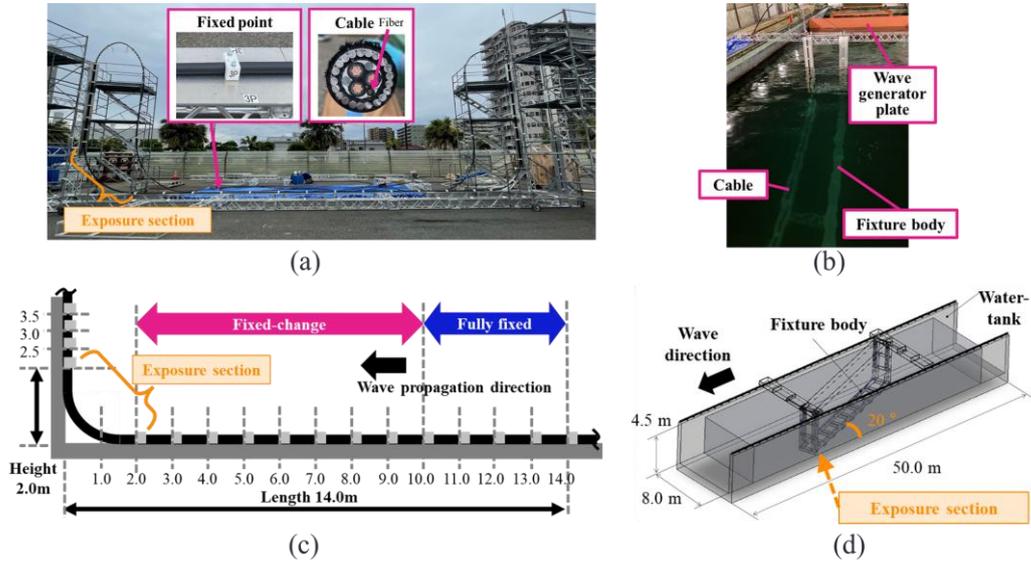

**Figure 3.** Experimental configuration of the free-span cable setup and wave-tank installation. (a) Overall view of the cable fixation structure assembled on land. (b) View from above of the submerged cable fixture in the wave tank. (c) Schematic illustration of the cable fixation and free-span configuration with adjustable unsupported length (2–10 m). (d) Installation of the cable fixture in the wave tank equipped with a wave generator.

The experimental conditions for DAS measurement and wave generation are summarized in Table 1. The wave heights and periods were selected considering the structural and operational constraints of the wave tank and wave generator. Although wave height and period are expected to influence the hydrodynamic loading and vibration response of the cable, the present study did not aim to establish their quantitative relationship with exposure length. Instead, limited wave conditions were used to verify the feasibility of the proposed method under controlled excitation. Wave height and period were varied to represent different excitation scenarios, and each condition was tested for approximately 2 min with two to three repeated trials.

**Table 1.** Experimental conditions for DAS measurement and wave generation.

| Category | Parameter | Value |
|---|---|---|
| Wave conditions | Wave height | 15 cm, 30 cm |
|  | Wave period | 1.25 s, 2.5 s |
| DAS measurement | Optical pulse width | 20 ns |
|  | Gauge length | 1.6 m |
|  | Spatial sampling interval | 0.8 m |
|  | Sampling frequency | 2000 Hz |
| Trials | Duration per trial | ~2 min |
|  | Number of trials | 2–3 |

4.2 Conditions for signal processing in the application of the proposed method

In this study, the two cores of the optical fiber cable were treated as independent free-span sections, hereafter referred to as Section 1 (Core 1) and Section 2 (Core 2). Note that due to various factors, including distance from the DAS device, slight differences in the frequency characteristics of the signals measured in both sections were observed.

As a preprocessing step, the STFT was applied along the temporal axis to DAS signals measured at multiple distance channels within a 12 m segment. This segment extended from the starting point of the exposed section toward the seabed and included the entire exposed section. The STFT was computed using a 50-second window and a 5-second shift interval. From the resulting time–frequency representations, only frequency components below 4 Hz were retained to focus on low-frequency structural vibration behaviours. For each time window, frequency components from multiple spatial locations were concatenated along the distance dimension, forming a two-dimensional frequency × distance feature representation. These features were arranged into sample-wise input matrices and used as inputs to the regression model.

To evaluate performance, a holdout validation scheme was used to clearly separate the training and evaluation data. Among experimental cases with identical wave excitation, exposure length, and free-span section, the second trial was used for training, while the remaining trials were reserved for evaluation. During regression-model training, pairs of free-span sections with different exposure lengths were selected to learn features sensitive to exposure-length differences. Subsequently, independent one-class SVM models were trained for each section using latent features extracted under the initial exposure condition.

For anomaly-detection evaluation, only the same free-span sections used in training were considered. Evaluation data covering all exposure-length conditions were input into the trained models, and anomaly scores were computed for each trial. Anomaly-detection results are obtained by binarizing the anomaly scores relative to a baseline of zero. Let $L_0$ denote the initial exposure length used for anomaly-model training and $L_e$ denote the evaluation exposure length. *The exposure-length change* is defined as

$$\Delta L = L_e - L_0. \quad (7)$$

4.3 Evaluation Results

The proposed framework's anomaly-detection accuracy was initially evaluated. Figure 4 shows the anomaly scores for all evaluation trials, aggregated by exposure-length change. The horizontal axis represents the exposure-length change, calculated using Equation (7). Non-zero values of the exposure-length change indicate abnormal conditions: positive values represent an exposure-length extension, and negative values indicate shortening. The vertical axis shows the anomaly score, defined as the signed distance from the decision boundary of the single-class SVM, computed using Equation (6).

As shown in Fig. 4, the anomaly score decreases approximately monotonically with increasing absolute exposure-length change ($|\Delta L|$), indicating a systematic trend in the latent feature space. A strong correlation was observed between $|\Delta L|$ and the anomaly score ($r = -0.83$), suggesting that deviations from normal cable conditions become progressively more pronounced with increasing exposure length. The separation between normal conditions ($\Delta L = 0$) and abnormal conditions ($|\Delta L| \geq 1$ m) was quantified. Statistical significance was confirmed using the Mann–Whitney U test [13], with Holm-corrected p-values below $10^{-10}$ for all deviation lengths. The corresponding effect sizes ranged from moderate to large (Cliff's $\delta = 0.32$–$0.57$), indicating that the observed differences are statistically robust and practically meaningful.

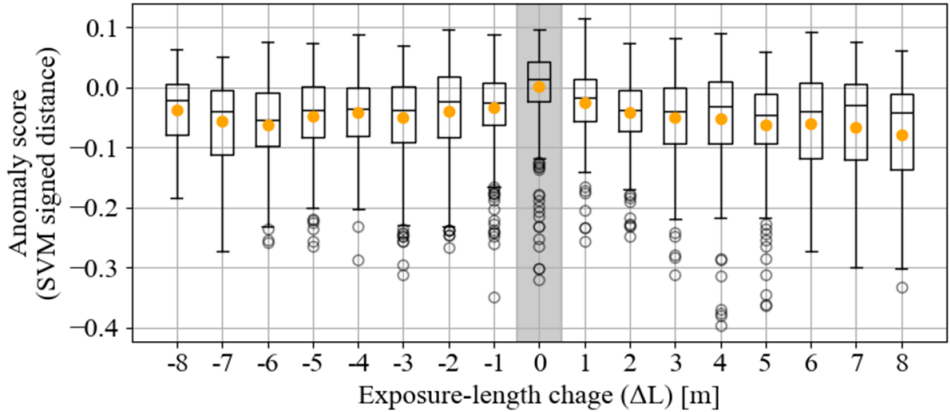

**Figure 4.** Anomaly scores for all evaluation trials, aggregated by exposure-length change (ΔL). The boxplots show the distribution of anomaly scores for each ΔL; hollow dots indicate outliers, and orange dots represent the mean. The anomaly score is defined as the signed distance from the decision boundary of the one-class SVM. Positive and negative ΔL values represent exposure-length extension and shortening, respectively.

To evaluate practical detectability, anomaly scores were further binarized to classify normal and abnormal conditions. Despite being trained on only 2 minutes of normal-condition data from a single trial, the proposed method achieved an accuracy of 0.73 and an F1 score of 0.82. These results confirm that exposure-length variations can be reliably detected even with minimal training data. Consequently, the framework satisfies a key operational requirement for offshore cable monitoring: enabling early-stage deployment immediately after installation without extensive abnormal data collection.

To further investigate whether the low-dimensional latent features extracted in the feature extraction step retain information about exposure length, the exposure length of the evaluation data was estimated using the regression model used for feature extraction. Figure 5 shows the relationship between the true and estimated exposure lengths. A clear positive linear relationship is observed, with a correlation coefficient of 0.72. The mean absolute estimation error was 1.20 m.

It should be emphasized that the purpose of the regression model in the proposed framework is not to provide precise estimates of exposure length. Instead, the regression model learns latent representations that capture exposure-length-dependent characteristics of cable vibration while suppressing environmental variability. From this perspective, the observed correlation between the estimated and true exposure lengths confirms that the latent features preserve information relevant to exposure-length differences, which is sufficient for the subsequent anomaly detection task.

Taken together, the strong correlation between the exposure-length change and the anomaly scores, along with the positive relationship between the ground-truth exposure length and the estimated exposure length, indicate that the proposed feature extraction method captures exposure-length-related characteristics of submarine cable vibrations. These results support the feasibility of detecting exposure-length variations under non-stationary environmental conditions with limited training data in a controlled laboratory setting.

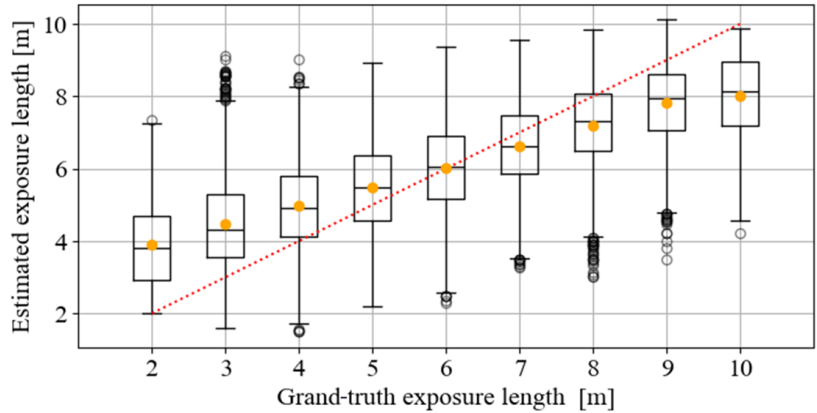

**Figure 5.** True exposure length versus regression-estimated exposure length for the evaluation data. The boxplots show the distribution of estimated exposure length for each true exposure length; hollow dots indicate outliers, and orange dots indicate the mean. The red dotted line represents the identity line (y = x). A positive linear relationship is observed (r = 0.72), with a mean absolute error of 1.20 m.

## 5. Conclusions

This paper proposed an anomaly detection method based on intermediate representations learned by a regression model using DAS to monitor exposure-length variations of free-span sections in submarine power cables for offshore wind farms. Wave-tank experiments under controlled conditions showed that the anomaly score tended to decrease monotonically with increasing exposure-length change (r = –0.83), and statistical testing confirmed significant separation between normal and abnormal conditions. Despite being trained with only one normal-condition dataset, the method achieved an accuracy of 0.73 and an F1 score of 0.82.

These results provide an initial laboratory-scale validation of the proposed method and suggest its feasibility for detecting exposure-length variations from DAS measurements. However, the present experiments were conducted under limited and simplified conditions and do not fully represent the complexity of real offshore environments. Therefore, further validation under actual offshore conditions is necessary to assess practical applicability and robustness against environmental variability. Through such stepwise validation, the proposed method is expected to contribute to the realization of Condition-Based Maintenance (CBM) for submarine power cables and to improve the long-term reliability and stable operation of offshore wind power systems.

## Acknowledgments

The experiments conducted in this study used the wave tank at the National Maritime Research Institute, National Institute of Maritime, Port and Aviation Technology (NMRI, NIMPAT), as well as power cables manufactured by OCC Corporation. In addition, we received significant support and cooperation in the preparation and execution of the experiments from Marine Works Japan Ltd., Ocean Works Asia Inc., and NEC Networks & System Integration Corporation. The authors gratefully acknowledge all of their contributions.